# THE LIPID-RNA WORLD

Saurav Mallik and Sudip Kundu*

Department of Biophysics, Molecular Biology and Bioinformatics, University of Calcutta; Address: 92, APC Road, Kolkata; India; Zip: 700009

"The only way of discovering the limits of the possible is to venture a little way past them into the impossible."

- Arthur C. Clarke (*Profiles of the Future*, 1962)

**ABSTRACT**

The simplest possible beginning of abiogenesis has been a riddle from the last century, which is most successfully solved by the Lipid World hypothesis. However, origin of the next stages of evolution starting form lipids is still in dark. We propose a 'Lipid-RNA World Scenario' based on the assumption that modern stable lipid-RNA interactions are molecular fossils of an ancient stage of evolution when RNA World originated from Lipid World. In accordance to the faint young sun conditions, we present an 'ice-covered hydrothermal vent' model of Hadean Ocean. Our hypothetical model suggests that faint young sun condition probably provided susceptible physical conditions for an evolutionary route from Lipid-World to Protein-RNA World, through an intermediate Lipid-RNA World. Ancient ribozymes were 'protected' by lipids assuring their survival in prebiotic ocean. The origin of natural selection ensures transition of Lipid-RNA World to Protein-RNA World after the origin of ribosome. Assuming the modern peptidyl-transferase as the proto-ribosome, we have presented a hypothetical translation mechanism: proto-ribosome randomly polymerized amino acids being attached to the inner layer of a lipid-vesicle, using only physical energies available from our Hadean Ocean model. In accordance to the strategy of chemical evolution, we also have described the possible evolutionary behavior of this proto-ribosome, which explains the contemporary three-dimensional structure of 50S subunit and supports the previous predictions regarding the ancient regions of it. It also explains the origin of membrane-free 'minimal ribosome' in the time of Last Universal Common Ancestor.

## 1 INTRODUCTION

### 1.1 THE RNA WORLD HYPOTHESIS

The possible mechanism of early evolution and origin of life on earth has been a central problem for evolutionists for centuries. The modern form of life operates by means of sequence specificity, which is responsible for structural diversity of biomolecules. Various structures are assigned with various enzymatic functions and heredity. Thus, the sequence specificity seems to be the fundamental property of life. Due to the detection of enzymatic function (1-6) of RNAs and presence of them as genetic materials in some virus, and its self-replicating property (1, 2, 7), RNA was hypothesized to initiate the earliest form of life. This led to an increased interest in the hypothesis that an RNA World, a term introduced by Gilbert (8), preceded the DNA/RNA/Protein world

*To whom correspondence should be addressed.
Email: saurav.bmb@gmail.com or skbmbg@caluniv.ac.in

(9). A detailed discussion of this hypothesis can be found in Supplementary Materials section 1.1.

In spite of its popularity, the RNA World hypothesis had to face strong arguments against its possible existence. Firstly, RNA oligomers itself seems unlikely to have emerged spontaneously in a primordial environment (9, 10). Secondly, according to geological researches (11-13), in a hot, acidic prebiotic environment RNA breaks upon hydrolysis (14). Fourthly, primitive earth was completely exposed to UV and gamma ray radiations (no ozone layer). The aromatic bases of RNA absorb strongly in the ultraviolet region, which makes them more susceptible to damage by background radiation (15). Fifthly, catalytic activities of RNAs are very limited (16) and are properties of relatively long RNA chains (17). We have presented a detailed critique on the RNA World in Supplementary Materials section 1.2.

### 1.2 ALTERNATIVE HYPOTHESIS: THE LIPID WORLD

On this context, some alternative hypotheses were proposed: for example 'proteins first' (18-23), virus world (24), lipid world (25) and prion-RNA world (15). However, except the lipid World hypothesis, all the others are inadequate in the sense that they claim the origin of functional proteins prior to the origin of a functional ribosome. This seems unlikely. Only the Lipid World hypothesis probably provides the most intuitive description of the possibilities in Hadean Eon. Scenarios in which replication is a collective property of a loose molecular assembly could be as likely to reflect the earliest stages of abiogenesis (26-29). The notion that lipids and other amphiphiles could serve as intermediates in prebiotic evolution has also been specifically elaborated (30-32). It has been proposed that lipid membranes may have a hereditary potential, as most membranes are generated from other membranes but not created de novo (33, 34). Lipid World hypothesis proposes the existence of self-aggregating lipid vesicles that have the heredity potential in terms of generating 'daughter' vesicles through fission and fusion interactions. A capacity for high fidelity transfer of compositional information could be acquired gradually, after many growth and division cycles, leading to a process akin to self-reproduction (35). Catalytic activities of lipids have also been proposed (36).

## 2 THE BEGINNING OF ABIOGENESIS

### 2.1 THE HADEAN OCEAN ENVIRONMENT

The Study of zircons has found that liquid hydrosphere (and so the atmosphere) must have existed as long ago as 4.4 Gyr, very soon after the formation of the Earth (37). Several hypothesized models show that the early evolution of life was linked to ancient hydrothermal activities in the Hadean and the early Archean (38-40). On the other hand, following the faint young sun paradox (41) a number of models were proposed proposing about a cool early earth (37, 42). A diplomatic solution for the two contradictory hypothe-sis was proposed by Bada *et al.* (43) in 1994 and Cleaves *et al.* in 1998 (44) that Hadean ocean were frozen only at the top, having approximately 300-1000 meter thick ice layer, which protected the early biomolecules from UV and gamma ray radiation. This model also provides a cold environmental situation where an RNA World can emerge and evolve.

Reviewing these proposals, we propose an 'ice-covered hydrothermal vent' model of Hadean Ocean. This is a simple modification of a common hydrothermal vent system with a frozen top. Submarine volcanoes establish a biologically rich alkaline (pH 9-10) high temperature (~350°C) sea floor, where the ocean top is frozen due to the faint young sun. Distant regions from volcanoes are acidic (pH 4-5). The hot water of sea floor will continuously experience an upstream motion towards the top, where it will lose temperature due to the ice layer and again come back to the sea floor. This will give rise to a continuous circular current of water from top to bottom. This continuous current will protect the organic materials from UV by taking them deep into water repeatedly. In addition, the ice layer at the top surface of ocean provides an extra protection. A detailed discussion about this model is included in supplementary materials section 2.1 and Figure-S1 schematically represents this scenario.

**2.2 ORIGIN OF NATURAL SELECTION**

Biological evolution is a gradual complexation of the living systems, by 'fine tuning' the enzymatic functions and heredity mechanisms through continuous cycles of self-reproduction at a large timescale. Abiogenesis started with a Lipid World, when the diverse biochemical reaction network (property of cellular life) was absent. If there was no cellular life, there should not be any evolutionary pressure on maintaining any biochemical reaction. Thus, in this scenario, the natural selection of mutations could not be processed in the contemporary way. The strategy of natural selection should be dependent and consistent with the physical conditions of the habitat of the biomolecules (consistency with habitat is the basic feature of natural selection).

An ice covered hydrothermal vent system with water temperature ranging from ~350°C at the bottom and ~0°C at the top and a strong pH gradient required structural consistency of the earliest biomolecules. Otherwise, they would have denatured soon in this hostile condition. In fact, the evolutionary strategy of biomolecules should be solely dependent on their structural stability. If a new monomer joins a polymer structure at a certain stage, its persistency will be determined by the structural stability it is providing to the polymer. If the monomer destabilizes the polymer, it should be dissociated soon. On the other hand, if it provides structural stability, its persistency will be high. Same kind of natural selection strategy should also be observed in case of isoenzymes: between two isoenzymes, the one with higher structural stability and higher catalytic activity should undergo natural selection. Thus, during abiogenesis, only those chemical reactions will undergo natural selection, whose components are structurally consistent and product yields are high. Eventually such reactions will organize and biochemical pathways will evolve. Therefore, we can identify a Darwinian picture in chemical evolution. We shall show later that this hypothetical strategy of natural selection fits perfectly with the evolutionary picture of ribosome, a remnant of RNA World. A detailed discussion on this topic is included in supplementary materials section 2.2.

**3 CONSTRUCTING THE IDEA OF A LIPID-RNA WORLD**

**3.1 LIPIDS IN PREBIOTIC WORLD**

Bangham *et al.* (45) first reported that phospholipids have the capacity to self-assemble. It was found that a variety of single-chain amphiphiles could self-assemble into bilayers under certain conditions. Ourisson and Nakatani (31) and others (46, 47) showed that isopentenol and dipolyprenol phosphates, either individually or as mixtures, are able to produce stable bilayer vesicles and membranes. Some other interesting studies have shown that when fatty acid micelles are added to a buffered solution containing a dispersion of particles (clay mineral), vesicles are produced at an accelerated rate, the particles become coated with amphiphiles, and particles are often encapsulated in the vesicles (45, 48). A more recent study (49) experimentally demonstrates that a microcapillary thermal diffusion column can concentrate dilute solutions of nucleotides, oligonucleotides, and fatty acids. Upon concentration, the self-assembly of large vesicles containing encapsulated DNA occurs in regions where the critical aggregate concentration of the fatty acid is exceeded.

The common presence of terpenoid and hopanoid derivatives as molecular fossils (31) show direct evidence that they were present in prebiotic world and were utilized by microorganisms. It is well known that such compounds, either individually or in mixture, are able to produce bilayers (46, 47). Deamer and his coworkers (50) observed many different structures in Murchison meteorite, most interestingly the membranous vesicles obtained with some but not all of the separated components. In a few cases, vesicles were observed that seemed to consist of an interior compartment surrounded by a double membrane. Dworkin and coworkers have demonstrated very similar results with amphyphylic material obtained in simulations of organic synthesis in interstellar ices subjected to UV irradiation (51). Self-assembling lipids can be synthesized in simulated interstellar/precometary ices, as described by Deamer *et al.* (50). These studies suggest that the prebiotic formation of vesicles (and even vesicle within a vesicle) was possible. A detailed picture of this section can be found in reference 25.

**3.2 VESICLE ENCAPSULATION AND ITS BIOLOGICAL ADVANTAGES IN PREBIOTIC WORLD**

The ice covered hydrothermal vent model enables us to observe the Hadean Ocean as a large thermal column with circulating water. The biologically rich environment and pH gradient is a susceptible condition for formation of a large number of lipid vesicles. Our ocean model represents a situation analogous to the experimental conditions described by Hargreaves *et al.* (48) and Budin *et al.* (49). Presence of montmorillonite in prebiotic ocean (10) would enhance the lipid aggregation rate, generating an extensive amount of vesicles encapsulating montmorillonite and other organic materials inside. Segré *et al.* (25) in their Lipid-World hypothesis described such an encapsulation event. The biological advantages of the prebiotic biomolecules of being vesicle encapsulated are pointed out in the following.

**3.2.1    Local phases of biologically rich environment:** Earlier works (52, 53) calculated the concentration of organic compounds in Archean seawater and ended up with a conclusion that they were almost similar to modern seawater concentrations. Therefore, in the acidic primitive ocean, the concentrations of organic materials were potentially too low for conducting covalent chemical reactions (25). On this aspect, Segré *et al.* (25) described one important advantage of vesicle encapsulation that it generates local phases of biologically rich environment (45, 48, 49), which means high concentration of organic materials inside. Such vesicle-encapsulated phases are more capable for biochemical reactions forming covalent bonds.

**3.2.2    Polymerization of nucleic acid and amino acids inside vesicles:** The formation of vesicles in a hydrothermal column causes encapsulation of montmorillonite, amino- and nucleic acids (see Supplementary Materials section 1.2.2-1.2.3). The montmorillonite is known to catalyze nucleic acid polymerization (54, 55). Tsukahara *et al.* in 2002 (56) experimentally showed that amino acids are polymerized on or inside lipid vesicles if they are subjected to the circulating water current in a hydrothermal vent with top and bottom temperatures being 0° and 180°C respectively. Thus, the hydrothermal vent model enables the Lipid World to synthesize peptide oligomers (even before the origin of ribosome). We shall show later that this lipid-mediated oligopeptide formation had a very important role in proto-ribosome function.

**3.2.3    Catalytic activity of Lipids:** Lipid assemblies are known to have some catalytic activity. They can enhance the rates of some chemical reactions (33, 57). Lipid vesicles potentially enhance rates of many biochemical reactions (58, 59). This rate enhancement can even lead to autocatalytic growth of such vesicles/assemblies (60). These catalytic functions were later taken over by ribozymes when an RNA World established.

**3.2.4    Vesicle encapsulation protects from UV:** Vesicle encapsulation can protect the fragile prebiotic RNA segments from UV (a strong argument against RNA World hypothesis) and other radioactive radiations. Once RNA enters inside a vesicle (mechanism will be described later), it is somewhat protected from the cosmic radiations of early earth (UV and Gamma), since lipids absorb such radiations (61, 62). Lipid peroxidation due to UV radiation might be delayed in our 'ice-covered hydrothermal vent' model of Hadean Ocean. The thin ice covering at the top of ocean provides initial protection. The continuous circular flow of water (along with organic materials) from top to bottom can protect the lipid vesicle from peroxidation and provide a long lifetime.

**3.2.5    Lipids as stabilizing agents:** Lipids can even work as stabilizing agents; proteins and RNAs are commonly expressed on membranes and liposomes. Membrane interactions with DNA/RNA and stability analysis of such complexes have been a matter of good research interest because of the use of liposomes as vectors of gene delivery (63, 64). The history of stability analysis of DNA-lipid complexes is quite old (65-71). Examples of lipid-RNA binding are not that diverse as DNA (72-76). However, Tajmir-Riahi *et al.* (77) have reported that in case of lipid-tRNA binding, RNA have a higher binding constant for membrane (i.e. more stable association) than those of the lipid–DNA complexes (78). This indefinitely states that even in a prebiotic world, there was a high possibility that the structural stability to the ancient ribozymes were provided by vesicle membranes they were encapsulated.

**3.3    MODERN LIPID-RNA INTERACTIONS**

RNA can bind to both interior (73, 79) and exterior (78) of liposomes. Exterior RNA binding with liposome membrane is linear over 1000-fold range of RNA concentrations (73). RNAs that can bind, change the permeability, and disrupt phospholipid bilayers have been isolated using selection–amplification (73, 74). The oligomerizing heterotrimer RNA(9:9:10)$_n$ (Fig. 1, ref. 73) was one particular example, found to bind stably to the exposed face of a phosphatidylcholine vesicle. Many RNA complexes with similar activities were isolated, with no sequences in common. It is observed that many RNA structures exist that bind phospholipid membranes (80). Membrane affinity is, in this sense, a simple molecular property, so it is likely that mutation would continuously produce such RNAs (80). This argument could be a rationale of the origin of ancient membrane-bound RNAs in prebiotic earth. Moreover, it has also been shown that RNAs can bind to lipid vesicles and alter the membrane permeability (both positive and negative regulation, ref. 74). Such membrane permeability alteration can take place in primitive vesicles also, allowing inflow of small charged biological molecules and metal ions into the vesicle from water body, thus creating an ion gradient across the membrane.

RNAs normally bind to the polar side of the membrane, probably through Hydrogen bonds and van der Waals contacts. RNA 9, RNA 10, RNA 67-2, RNA 10$_{TRP}$, RNA 80N, RNA 10Arg(D)5, and RNA Trp70-93 are established membrane-interacting ribonucleotide polymers of ~100-residue length (72). Janas *et al.* (72) have shown that a greater fraction of RNA structures bind bilayers as phospholipid order increases: fluid<raft<ripple<gel. We shall show that the proto ribosome once polymerized amino acids based on this molecular property. RNA binding to lipid ripple-gel phase (at the phase transition temperature) unexpectedly shows no significant RNA-structure specificity. On the other hand, a state of intermediate order (the liquid-ordered or rafted state, believed to occur in biological membranes) shows RNA-structure dependent affinity. To account for this, Khvorova *et al.* in 1999 (74) and Janas *et al.* in 2006 (72) predicted secondary structures of the membrane-interacting RNA sequences. All the predicted secondary structures contained one or more helical segments and, in a few cases, a centroid region (Table-1, ref. 68 and Figure-4, ref. 74). In addition, it is obvious from the predicted secondary structures that the three-dimensional structure of these RNA segments must be planner, which can 'fit' on the membrane surface and cause binding. On this context, we argue that any ancient ribozyme having a planner three-dimensional structure and having a centroid and/or at least one linear helical secondary structural segment could possibly have a membrane-bound origin.

In the next sections, we shall show that predicted proto-ribosome structure fits excellently with these structural requirements. In fact, this discussion, in accordance with our hypothesis, can lead to an intuitive argument that modern RNA-lipid complexes are molecular fossils of an ancient stage of evolution when lipids provided

RNAs structural stability essential to survive in prebiotic environment. In the hot primitive ocean, it is quite intuitive to imagine the mutual stability of ribozymes interacting with lipid vesicles. Such a stabilized assembly should be more susceptible for a directional evolution compared to independent presence of RNAs and lipids, in a Darwinian picture. In addition, such assemblies can result in the ribozyme take over of lipid-catalytic activities and can initiate the transition of Lipid World to a Lipid-RNA World.

### 3.4 EVOLUTIONARY ROUTE TOWARDS A LIPID-RNA WORLD

The transition of Lipid World to the next stages of evolution should have occurred by means of Lipid-mediated polymerization of biomolecules. Segré *et al.* provided a hypothesis about the possible mechanism of polypeptide and polynucleotide synthesis from lipid vesicles/membranes (see ref. 25). However, there are possible alternative mechanisms of polymerization of nucleotides. Hanczyc *et al.* in 2003 (81) reported that montmorillonite catalyze the formation of closed vesicles from micelles composed of simple aliphatic carboxylic acids and that particles of the clay become encapsulated within the vesicles. Since montmorillonite are excellent catalysts for the oligomerization of a number of activated nucleotides, this might point to a route to a nucleic acid synthesizing system enclosed within a vesicle (10). We have already discussed that the thermal diffusion column-like behavior of Hadean ocean might provide a susceptible environment for encapsulation of both montmorillonite and organic materials.

The importance of montmorillonite in RNA structures was discussed in detail by Pyle in 2002 (82). Steitz and coworkers in 2004 (83) reported the abundance and structural significance of cations in stabilizing 23S rRNA structure. They identified 116 $Mg^{2+}$ ions and 88 monovalent cations ($K^+$, $Na^+$) bound principally to 23S rRNA. Both divalent and monovalent cations stabilize the tertiary structure of 23S rRNA by mediating interactions between its structural domains. Bound metal ions are particularly abundant in the region surrounding the peptidyl transferase center (PTC), where stabilizing ribosomal proteins are notably absent. They proposed that this might point to the importance of metal ions for the stabilization of specific RNA structures in the evolutionary period prior to the appearance of proteins. In fact, Fox presented a number of logical arguments (84) to establish that PTC originated in a protein-free world. After vesicular intake of RNAs attached with montmorillonite occurred in prebiotic environment, ribonucleotides were further polymerized within vesicles, with polymerization rate and stabilization enhanced by both metallic ions and lipids.

### 4 THE ORIGIN OF RIBOSOME

So far, we have discussed in detail that prebiotic polymerization and stabilization of RNAs by lipids and metallic cations are possible. Next, ribozyme actions should lead to amino acid polymerization, so that with course of time, a protein-RNA World originates. However, a high rate of amino acid polymerization seems to be impossible without ribosome. Thus, we need to understand the origin and evolution of ribosome in a Lipid-RNA World. A detailed discussion about the origin of ribosome in accordance to several prior hypothetical pictures is presented in the Supplementary Materials section 3. In addition, we have presented there a critique on the previous predictions about the nature of translation machinery in a prebiotic environment. A conclusion of that discussion is demonstrated in the following.

### 4 FROM LIPID WORLD TO PROTEIN-RNA WORLD

The 'Primordial soup' theory (101-103) speculates about an ancient stage of evolution when organic materials were concentrated in prebiotic oceans (4.4 Gyr). Bernal suggested that there were a number of clearly defined 'stages' that could be recognized in explaining the origin of life (103): first stage, when biological monomers originated, second stage, when monomers were polymerized and the third stage, when polymers evolved into protocells. If compositional complexity of a system is visualized as the diversity of polymer structures and possible biochemical reactions, then the whole process can be visualized as a gradual increment of complexity.

From such a 'Primordial soup,' self-aggregation of lipids can form early membranes and vesicles. Experimental works of Hill and coworkers (104) shows that sufficiently long negatively charged oligomers (e.g., RNA) adsorb almost irreversibly on anion-exchanging minerals. Since template-directed synthesis occurs without interference on many mineral surfaces (105), mineral particles might support replication until they became saturated with the descendents of a single ancestral RNA molecule. We have already discussed that montmorillonite within minerals can catalyze ribonucleotide polymerization. Moreover, when lipid vesicles are formed in a thermal gradient solution containing mineral clay, those clay particles are often encapsulated by vesicles (49). This is a possible molecular process by which oligonucleotides were encapsulated inside lipid vesicles. Eventually, they might be incorporated on inner layer of vesicles, gaining extra stability by interacting with polar headgroups of lipid. Further polymerization of the RNA segments attached to metallic cations and lipids can originate a Lipid-RNA World. If the vesicles are exposed to an environment of fluctuating temperature, this will lead to rapid fission-fusion interaction (106, 107) between vesicles, which will eventually produce a vesicle pool where daughter vesicles are produced with high fidelity of compositional identity. This can be viewed as a somewhat complicated alternative of the 'evolution like behavior' described by Segré *et al.* (25) in their Lipid World hypothesis. Probably by this time, ribozymes that were capable of RNA-dependent RNA transcription evolved. This is essential for high fidelity heredity in a Lipid-RNA World. Nelson *et al.* (108) and Wochner *et al.* (109) described such ribozymes.

Thus, we have described the Lipid-RNA World as a large vesicle pool containing diverse types of ribozymes and the proto-ribosome, where vesicles are undergoing rapid budding and fission-fusion interactions upon temperature fluctuation. Synthesis of random polypeptides in a large vesicle pool maintains the potential of synthesizing many functional proteins, which could have played important roles in early evolution. Ribonucleoprotein (RNP) complexes originated by the random interactions between protein and RNA segments. As we described before, such complexes being structurally more stable than RNA-only structures, were subjected to natural selection in a Darwinian picture. This was the beginning of 'protein take over' of the ribozyme functions through intermediate RNP stages (110, 111).

One theoretical study by Atilgan and Sun (112) shows a critical concentration of conical membrane proteins or proteins with non-zero spontaneous curvature can drive the formation of small vesicles. The driving force of vesicle budding stems from the preference of proteins to gather in regions of high curvature. A sufficiently high concentration of proteins therefore can influence the topology of the membrane. In the limit of mechanical equilibrium with no thermal fluctuations, it was found that depending on their shapes, pairs of proteins could repel or attract each other with long-range forces (113-116). Moreover, protein interactions depend on the curvature of the membrane (116). All these lead to the hypothesis that proteins can induce membrane shape changes, budding and even fission-fusions. We suggest that these results entail that proteins could have played an important role in vesicle replication and establishment of a 'self-replicating organism' pool. Probably there was a transition phase in chemical evolution when vesicle fission-fusion interactions were turning to be more and more protein action-dependent and less temperature fluctuation dependent.

As the vesicles within a vesicle pool undergo self-replication processes, compositional informations are transferred to the next progeny. If the compositional complexity is slowly increasing, then after a large number of self-replicating cycles, a vesicle pool with high fidelity heredity will be established. However, as proteins were synthesized by the lipids and proto-ribosome, rate of complexity increment with time will be enhanced. Diversity of protein structures and functions might be the reason of it. Thus, as the Lipid-RNA World to Protein-RNA World transition occurs, high fidelity heredity is no longer possible. However, we can expect a balance between increasing complexity and high fidelity heredity in such transition phases. Proteins mediating vesicle budding and fission-fusion interactions might evolve as the cell-division related proteins in future and such vesicles can evolve as proto-cells. A schematic representation of lipid-RNA World scenario is schematically shown in Supplementary Materials Figure-S5 (A-F).

## SUMMARY

It is almost certain that self-replicating organism cannot appear in prebiotic earth from inorganic materials (rejecting the idea about extra-terrestrial origin of life) just by sudden. The idea of RNA World strongly requires this property. Hence, it is rational to imagine that chemical evolution had an even simpler beginning, which can eventually lead to self-replication. The Lipid World hypothesis deals with such a possibility. Assembly formation, catalytic activities and evolution-like behavior of lipids, along with the presence of lipid vesicles as molecular fossils strongly support this idea. RNA World seems to be the second stage of life when sequence specificity of biomolecules first ever coped with enzymatic function and heredity. It is an intuitive argument that proteins, constructed by twenty different types of components (amino acids) and having large structural variations are too complex to be the first enzymes during abiogenesis. On this context, we reject 'protein-first' hypothesis. On the other hand, RNA, in comparison to proteins has simpler monomers with less structural diversity that could originate well before proteins. Nevertheless, RNA is structurally fragile. Therefore, without the help of stabilizing agents it must face rapid degradation in hot acidic prebiotic oceans. Lipids and metallic ions could indefinitely serve this job as stabilizing agents. Hence, this logical picture enables us to redevelop the concepts of early evolution. In addition, we have tried to provide a molecular mechanism of protein synthesis in Lipid-RNA World, which would eventually take up the Lipid-RNA World to the next stage of chemical evolution: the Protein-RNA World. This model suggests that proto-ribosome could have polymerized amino acids without the need of any chemical energy source, just by utilizing physical forces generated by thermal fluctuations. Our ribosomal evolution model can explain the contemporary 23S rRNA structure and supports the previous predictions about its ancient regions. Synthesis of random polypeptides in a large vesicle pool maintains the potential of synthesizing many functional proteins, which could have played important roles in early evolution. It can be speculated that with course of time such vesicles evolved into proto-cells.

## ACKNOWLEDGEMENTS

The authors acknowledge the computational facility of DIC and Department of Biophysics, Molecular Biology and Bioinformatics, University of Calcutta to do the work. This work is not funded by any authority; the authors performed it on their personal interests.

## SUPPLEMENTARY MATERIALS

Supplementary Materials (Section 1-4, Figures S1-S5 and References 1-156) are available at PNAS online.